\documentstyle[12pt,aaspp4,epsf]{article}
\begin{document}

\title{OFF-CENTER MERGERS OF CLUSTERS OF GALAXIES AND 
       NONEQUIPARTITION OF ELECTRONS AND IONS IN INTRACLUSTER MEDIUM}
\author{{\sc Motokazu Takizawa}}
\affil{Department of Astronomy, Faculty of Science, Kyoto University, 
       Sakyo-ku, Kyoto 606-8502, JAPAN}
\affil{Research Center for the Early Universe, Graduate School of Science,
       University of Tokyo, Bunkyo-ku, Tokyo, 113-0033, Japan}
\authoremail{takizawa@astron.s.u-tokyo.ac.jp}

\begin{abstract}
We investigate the dynamical evolution of clusters of galaxies and
their observational consequences during 
off-center mergers, explicitly considering the relaxation process 
between ions and electrons in intracluster medium by N-body
and hydrodynamical simulations. In the contracting phase a bow shock is
formed between the two subclusters. The observed temperature between 
two peaks in this phase depends on the viewing angle even if the geometry
of the system seems to be very simple like head-on collisions.
Around the most contracting epoch, when we observe merging clusters
nearly along the collision axis, they look like spherical relaxed 
clusters with large temperature gradients.
In the expanding phase, spiral bow shocks occur. As in head-on
mergers, the electron temperature is significantly lower than the plasma
mean one especially in the post-shock regions in the expanding phase. 
When the systems have relatively large angular momentum, double-peak 
structures in the X-ray images can survive even after the most contracting 
epoch. Morphological features in both X-ray images and electron
temperature distribution
characteristic to off-center mergers are seriously affected by the viewing
angle. When the clusters are observed nearly along the collision axis, 
the distribution of galaxies' line-of-sight (LOS) velocities is a good 
indicator of mergers. In the contracting phase, an negative kurtosis 
and a large skewness are expected for nearly equal mass collisions and rather
different mass ones, respectively. To obtain statistically 
significant results, about $1000$ galaxies' LOS velocities are required.
For nearby clusters ($z<0.05$), large redshift surveys 
such as 2dF will enable us to study merger dynamics.
\end{abstract}


\section{INTRODUCTION}

According to the hierarchical clustering scenario such as the cold
dark matter cosmology, it is believed that clusters of galaxies 
(CGs) are formed through subcluster merger and/or absorption of smaller
groups. Mergers of CGs are the most energetic 
events in the Universe after the Big Bang, where the total kinetic energy
of the two subclusters reaches $10^{63-64}$ ergs. Thus, it is most likely 
that they affect various properties of CGs. In intracluster medium
(ICM), mergers make strong bulk-flow motion and shocks. These cause
characteristic morphology of X-ray images (elongation and/or substructures)
and complex electron temperature structures in ICM 
(e.g., Honda et al. 1996; Churazov et al. 1998; Donnelly et al. 1998; 
Donnelly et al. 1999; Markevitch, Sarazin, \& Vikhlinin 1998; 
Watanabe et al. 1999; Watanabe, Yamashita, \& Furuzawa 1999). 
In addition, rapid change of physical properties of ICM probably
induces nonequilibrium plasma ionization
(Hanami et al. 1999). For galaxies, mergers change drastically their
environments through gravitational tidal field (Bekki 1999) 
and static and ram pressure of ICM (Evrard 1991; Fujita et al. 1999). 
These may affect star formation activities of member galaxies
(e.g., Caldwell et al. 1993; Tomita et al. 1996; Wang, Ulmer, \& Lavery 
1997). Furthermore, high energy phenomena of CGs such as radio halos
(e.g., Giovannini et al. 1993; Deiss et al. 1997; R\"{o}ttgering et al. 
1997) and hard X-ray emission (Fusco-Femiano et al. 1999)
are well correlated to merger phenomena. This suggests a part of huge
energy of mergers is transported to acceleration of high energy 
relativistic particles (Sarazin 1999) and amplification of magnetic
field, though their detailed physical processes are still unclear.

Since CGs evolve in cosmological timescale and their initial conditions  
depend on the cosmology, cosmological N-body + 
hydrodynamical simulations are suitable to investigate the formation 
and evolution of CGs in general (e.g., Evrard 1990; 
Navarro, Frenk, \& White 1995; Bryan \& Norman 1998; 
Eke, Navarro, \& Frenk 1998; Suginohara \& Ostriker 1998;
Yoshikawa, Itoh, \& Suto 1998).
However, the simulations starting from rather idealized initial 
conditions are more suitable to investigate mergers in detail
(Roettiger, Burns, \& Loken 1996; Ishizaka \& Mineshige 1996).
Schindler \& M\"{u}ller (1993) found that characteristic temperature 
structures occur in merging clusters through shock heating and 
adiabatic compression and expansion. Ishizaka (1997) found that
the specific energy ratio of ICM and galaxies, $\beta_{\rm spec}$, 
is a good indicator of merging clusters and that it can be used to 
determine the phase of the mergers. Takizawa (1999) performed N-body
and hydrodynamical simulations explicitly considering the relaxation
process between ions and electrons and find that
electron temperature distribution becomes significantly different
than ion temperature one in later stages of mergers. 
Recently, Roettiger, Burns, \& Loken (1999)
performed N-body + magnetohydrodynamical simulations to investigate
evolution of magnetic field in ICM during mergers.

Some of the observational results (e.g., A754 in Henriksen \& Markevitch 
1996; A3395 in Markevitch et al. 1998) imply off-center mergers. 
Furthermore, cosmological
N-body simulations show collapsed dark halos have angular momentum
which correspond to $\lambda \sim 0.01 - 0.1$ (Barnes \& Efstathiou 1987; 
Ueda et al. 1994), where $\lambda$ is the ratio between the actual angular
velocity and the angular velocity needed to provide rotational support
(see Binney \& Tremaine 1987).
It is probable that dark halos obtain a part of the angular momentum
through off-center major mergers. Thus, it is an quite interesting problem
to study evolution of off-center mergers and their observational 
consequences. Ricker (1998) studies
gas dynamical evolution of off-center mergers in detail using the high 
resolution peacewise-parabolic method. However their simulations do
not contain the collisionless N-body component. This is very problematic
because dark matter and galaxies dominate gravity in typical CGs.
In addition, two-temperature nature of ICM is not considered there. 
Thus, we perform N-body and hydrodynamical simulations explicitly 
considering the relaxation process between 
ions and electrons to investigate the dynamical evolution and 
observational consequences of off-center mergers of CGs. 

The rest of this paper is organized as follows. In \S \ref{s:sim}
we describe the adopted numerical methods and initial conditions 
for our simulations. In \S \ref{s:res} we present the results. 
In \S \ref{s:sd} we summarize the results and discuss their implications.

\section{THE SIMULATIONS}\label{s:sim}

\subsection{The Numerical Method}

In the present study, we consider CGs consisting of two components:
collisionless particles corresponding to the galaxies and dark matter 
(DM), and two-temperature gas corresponding to the ICM. 
When calculating gravity, 
both components are considered, although the former dominates over the
latter. Radiative cooling and heat conduction are not included.
The numerical method used here is fully described in \S 3.1 of 
Takizawa (1999). Thus, we briefly review the outline of the method
in this subsection.

We used the smoothed-particle hydrodynamics (SPH) method to solve the
hydrodynamical equations for the gas component (see Monaghan 1992).
As the standard SPH codes for one-temperature fluid, we solve 
the continuity equation, the momentum equation, and the thermal energy
equation for whole plasma with artificial viscosity to treat the shocks. 
In addition to these equations, we solve one more equation for the 
normalized electron temperature, 
$\tilde{T}_{\rm e} \equiv T_{\rm e}/\bar{T}$, where 
$T_{\rm e}$ is the electron temperature and $\bar{T}$ is the plasma mean 
temperature. We assume that artificial viscous heating is effective only
for ions and that only the Coulomb coupling is considered in the 
relaxation process.

Gravitational forces are calculated by the Barnes-Hut tree algorithm
(Barnes \& Hut 1986) and softened using the Plummer potential profile. We 
set the softening length $\epsilon$ one-tenth of the initial core radius 
of the smaller subcluster in the simulation. Tree structure is also used 
to search for nearest neighbors in SPH calculations (Hernquist \& Katz 1989).

\subsection{Models and Initial Conditions}

We consider mergers of two virialized subclusters of galaxies
with masses $M_1$ and $M_2$. The initial configuration of each subcluster is 
the same as in Takizawa (1999). The spatial distribution of DM in each 
subcluster is represented by the King distribution with core radii 
$r_{\rm c,1}$ and $r_{\rm c, 2}$. We assume the velocity distribution of 
the DM particles to be an isotropic Maxwellian. We assume
that the initial ICM temperature is isothermal and equal to the virial 
temperature. The ICM is initially in hydrostatic equilibrium within the 
cluster potential of the DM and the ICM, itself.

We set the initial conditions as follows. Two subclusters are 
initialized in the $xy-$plane, separated by a distance $R$ in the
$x-$direction and a distance $b$ (the impact parameter) in the 
$y-$direction. The initial relative velocity, $v_{\rm init}$
is directed along the $x-$axis. The coordinate system is taken in such
a way that the center of masses is at rest in the origin.
The total gas mass fraction $f_{\rm g}$ is set to be 0.1. 

In both Run A1 and A2, two subclusters have the same masses; 
$M_1 = M_2 = 0.5 \times 10^{15} M_{\odot}$. We
set the core radii to be $r_{\rm c,1} = r_{\rm c,2} = 0.2$ Mpc. Each 
subcluster consists of 5000 collisionless particles and 5000 SPH particles.
For the both runs, the binding energy between the two subclusters is 
the same as in Run A of Takizawa (1999). We set $R = 3.2$ Mpc.
Then, the impact parameter $b$ and initial relative velocity $v$ are taken
in such a way that $\lambda = 0.01$ for Run A1 and $\lambda =0.02$ for
Run A2, respectively. The parameter $\lambda \equiv L |E|^{1/2}/(G M^{5/2})$ 
is the ratio between the actual angular velocity and the angular velocity 
needed to provide rotational support, where $L$ is the angular momentum 
of the two subclusters around the center of masses, $E$ is the binding energy 
between the two, $G$ is the gravitational constant, and $M$ is
the total mass (see Binney \& Tremaine 1987).

In both Run B1 and B2, the mass ratio is $M_1:M_2 = 4:1$. The larger 
subcluster has the same mass as that in Run A1 and A2. The particle numbers
of the smaller one are one-fourth of the larger one. 
For the both runs, the binding energy between the two subclusters is 
the same as in Run B of Takizawa (1999). We set $R = 2.2$ Mpc.
Again, the impact parameter $b$ and initial relative velocity $v$ are taken
in such a way that $\lambda = 0.01$ for Run B1 and $\lambda =0.02$ for
Run B2, respectively. 
The parameters in our calculations are summarized in Table \ref{tab:para}.

\section{RESULTS}\label{s:res}

\subsection{X-ray Images and Electron Temperature Distribution of Run A1}\label{s:xietmra1}

\begin{figure}
    \epsfxsize = 12 cm
    \centerline{\epsfbox{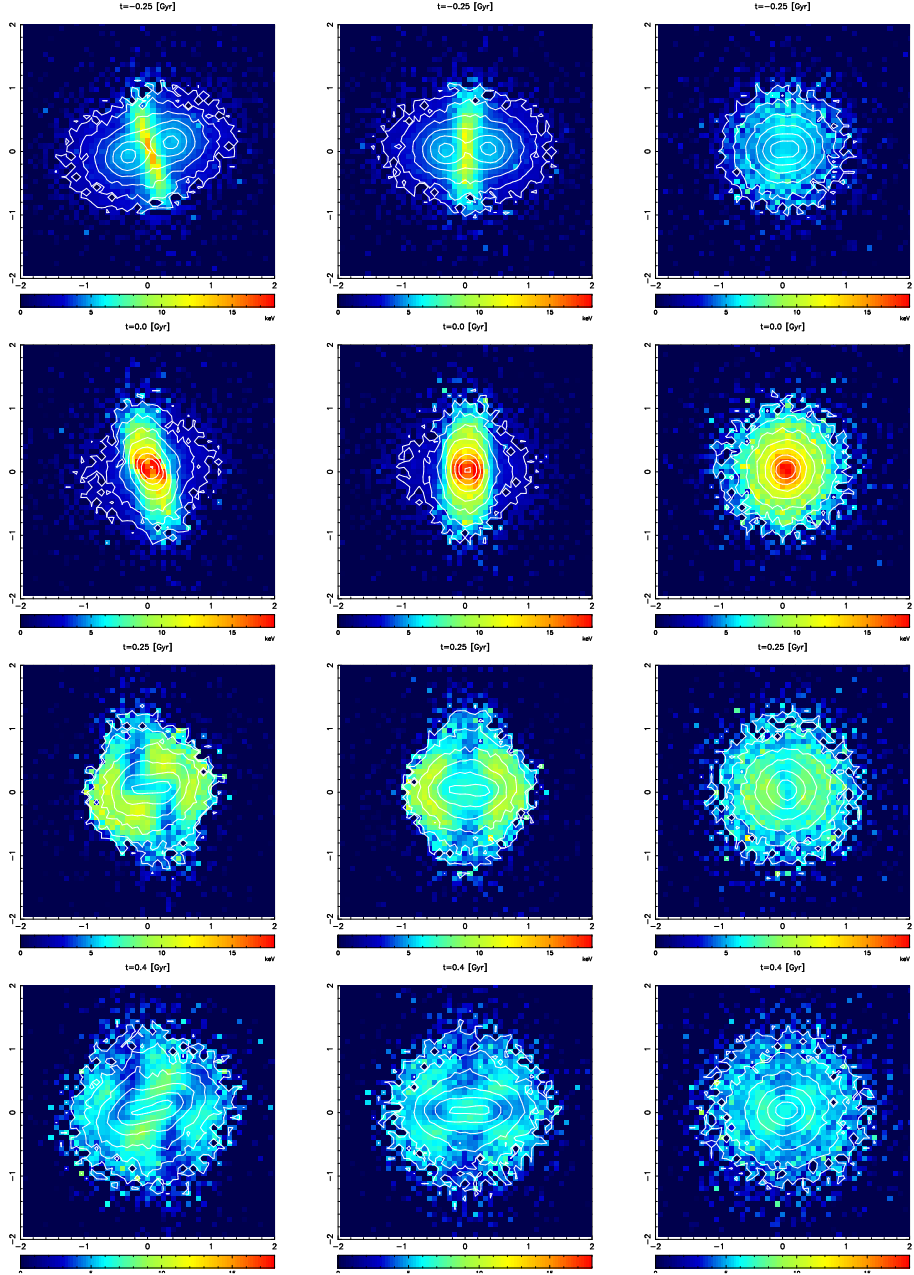}}
\caption{The snapshots of X-ray surface brightness (contours) and 
         emissivity-weighted electron temperature (colors) distribution of 
         Run A1 viewed along the $z$-axis (Column 1), $y$-axis (Column 2), 
         and $x$-axis (Column 3) at four epochs (Rows 1-4) during the merger. 
         The times relative to the most contracting epoch are listed above 
         each panel: $t=-0.25$, $0.0$, $0.25$, and $0.4$ Gyr. 
         X-ray surface brightness contours are equally spaced on a 
         logarithmic scale and separated by a factor of 7.4. The blue, 
         green, yellow, and red colors correspond to 
         $k_{\rm B} T \sim 4$ keV, 8 keV, 12 keV, and 16 keV, respectively.
         Each panel is 4 Mpc on a side.}
\label{fig:xrsbtea1}
\end{figure}

First of all, let us see the evolution of electron temperature 
($T_{\rm e}$ ) distribution and X-ray surface brightness distribution,
which are important observational quantities in X-ray to investigate 
the dynamical properties of merging CGs. Figure \ref{fig:xrsbtea1} 
shows the snapshots
of X-ray surface brightness (contours) and emissivity-weighted $T_{\rm e}$
(colors) distribution of Run A1 viewed along the $z$-axis (Column 1), 
$y$-axis (Column 2), and $x$-axis (Column 3) at four epochs (Rows 1-4) 
during the merger. The times relative to the most contracting epoch
are listed above each panel: $t=-0.25$, $0.0$, 
$0.25$, and $0.4$ Gyr. X-ray surface brightness contours are equally spaced 
on a logarithmic scale and separated by a factor of 7.4. The blue, green,
yellow, and red colors correspond to $k_{\rm B} T \sim 4$ keV, 8 keV, 
12 keV, and 16 keV, respectively. Each panel is 4 Mpc on a side.

When we observe the cluster along the z-axis (Column 1), 
which is nearly perpendicular to the collision axis, the morphology 
characteristic to the off-center merger is clarified. When two 
subclusters approach each other, we see double peaks in the X-ray 
image of ``one cluster'' at $t=-0.25$ Gyr. Just between the peaks 
electron temperature rises up to $\sim 15$ keV due to the shock. The shock
front is oblique to the collision axis due to the angular momentum between 
the subclusters. At the most contracting epoch ($t=0.0$ Gyr), 
two peaks merge to one peak in the X-ray image and the image
elongates to the directions nearly perpendicular to the collision
axis. Then the cluster expands and two shocks propagate in the opposite
directions ($t=0.25$ Gyr). Since the system has angular momentum, 
the shape of the expanding
shocks are similar to the arms of the spiral galaxies. However, due to the 
limitation of relaxation between ions and electrons, the post-shock hot 
regions have not so high electron temperature as in the contracting phase
(see \S \ref{s:tniep}).

On the other hand, when we observe the cluster along the y-axis (Column 2),
it looks like a head-on merger such as Run A
in Takizawa (1999). The $T_{\rm e}$ between the two peaks 
at $t=-0.25$ Gyr becomes lower than when viewed along the z-axis,
because the both foreground and background cooler gas contaminates the hot gas
at the shock. This lead to the underestimation of the collision velocity 
though the morphology of this epoch is very simple and seems to be suitable 
for the estimation (and we will discuss this in \S \ref{s:sd}).

Furthermore, when we observe the cluster along the x-axis (Column 3),
which is nearly along the collision axis, the morphology is rather different.
It looks like a spherical symmetric cluster both in the X-ray surface 
brightness and electron temperature distribution. Except for the most 
contracting epoch ($t=0.0$ Gyr), there is no distinct temperature
structure. At the most contracting epoch, temperature distribution is 
spherical symmetric but have a rather large gradient (and we will discuss 
this in \S \ref{s:sd}). Since ICM is optically thin, information along 
the line-of-sight direction is emissivity-weighted 
integrated and lost in large part. In this case, line-of-sight velocity
distribution of galaxies is more suitable for analyzing the cluster dynamical
properties (see \S \ref{s:lvdg}). In addition, line-of-sight velocity of ICM
may also provide us with useful information, which will become observable 
with the X-Ray Spectrometer (XRS) after the launch of ASTRO-E.

\subsection{X-ray Images and Electron Temperature Distribution of Run B1}\label{s:xietmrb1}

\begin{figure}
    \epsfxsize = 12 cm
    \centerline{\epsfbox{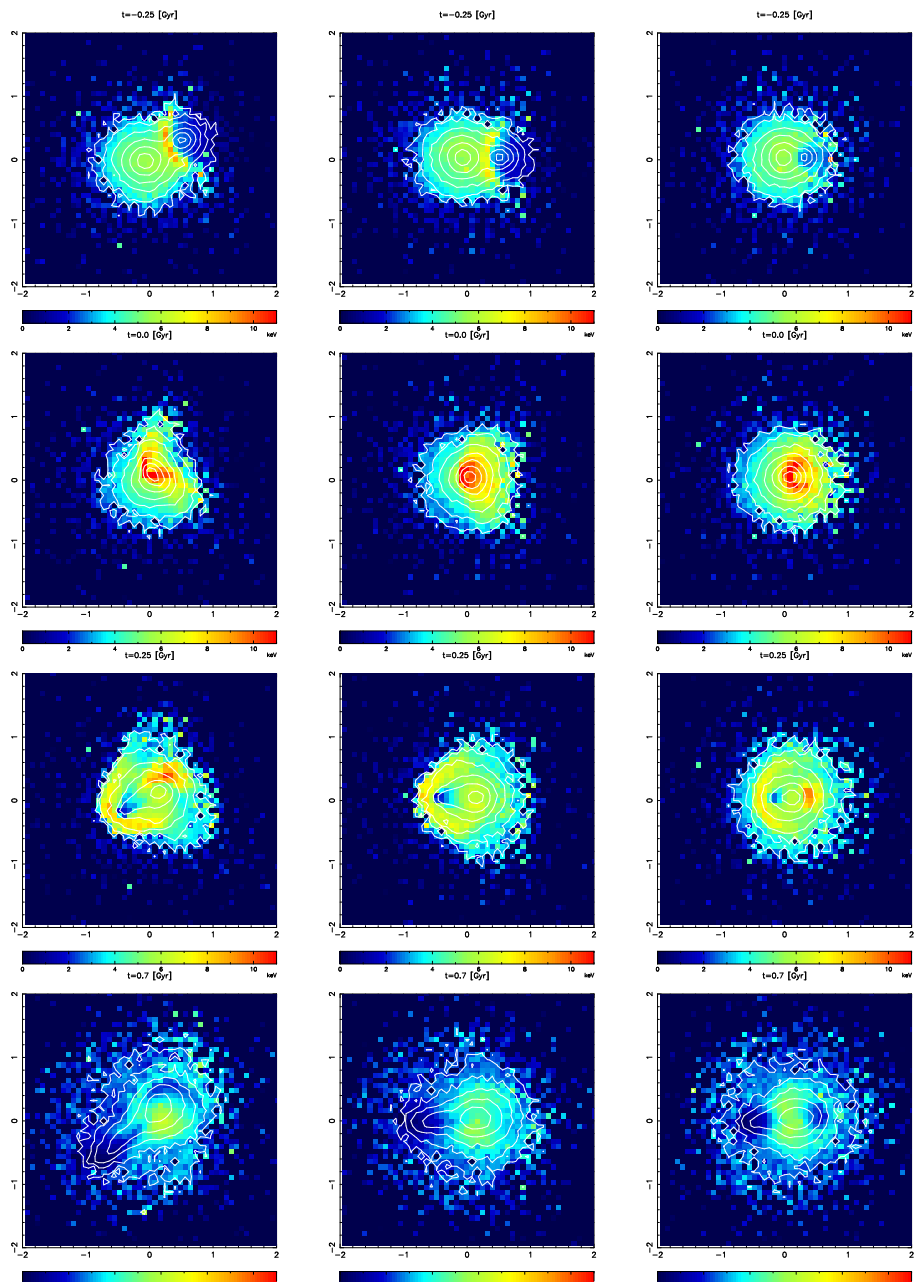}}
\caption{The same as figure \ref{fig:xrsbtea1}, but for Run B1. 
         The times are listed above each panel: 
         $t=-0.25$, 0.0, 0.25, and 0.7 Gyr. Temperature color scale is 
         adjusted for Run B1. The blue, green, yellow, and red colors 
         correspond to $k_{\rm B} T \sim 2.5$ keV, 5 keV, 7.5 keV, 
         and 10 keV, respectively. }
\label{fig:xrsbteb1}
\end{figure}

Next, we describe the evolution of X-ray surface brightness and $T_{\rm e}$ 
distribution of Run B1. Figure \ref{fig:xrsbteb1} shows the same conditions as
Figure \ref{fig:xrsbtea1}, but for Run B1. The times are listed above each 
panel: $t=-0.25$, $0.0$, $0.25$, and $0.7$ Gyr. Temperature color scale is 
adjusted for Run B1. The blue, green, yellow, and red colors correspond to 
$k_{\rm B} T \sim 2.5$ keV, 5 keV, 7.5 keV, and 10 keV, respectively.

As in Run A1, characteristic features of off-center mergers
are most clearly seen when the cluster is viewed along the z-axis
(Column 1). When two subclusters approach each other, the bow shock with an 
arc shape is formed just between the two as in head-on collisions. 
However, since the direction of 
the smaller subcluster's velocity is not coincident to the direction 
connecting the two, the shock is stronger in the upper region of the 
panel ($t=-0.25$ Gyr). Then the two peaks merge to one triangle image 
($t=0.0$ Gyr). The hot region associated to the bow shock is seen 
elongated slantingly backward with respect to the motion of the smaller
subcluster. However, this hot region is located not in the center of the
cluster but in the upper half of the image due to an off-center collision.
Then the gas expands and 
the two shocks propagate outward ($t=0.25$ Gyr). 
As in Run A1, the expanding 
shocks form spiral-like structure and electron temperature there is 
significantly lower than the plasma mean temperature 
(see \S \ref{s:tniep}).

When we observe the cluster along the y-axis (Column 2), its morphological
feature in X-ray surface brightness is similar to the that of head-on merger
as in Run A1. However, this is not the case in the electron
temperature distribution. This is prominent especially in the expanding 
phase ($t=0.25$ Gyr). The hot region located in the backward with
respect to the motion of the smaller cluster is hardly seen. Since this 
hot component is not so spread, the foreground and
background cooler gas contaminates it. When we observe the cluster along 
the x-axis (Column 3), non spherical structure can be seen 
on the contrary to the Run A1. At $t=-0.25$ the geometry is very simple
and seems to be very suitable for estimation of the collision velocity.
However, this lead to the underestimation of the collision velocity 
as in Run A1.

\subsection{Two-Temperature Nature of ICM in the Expanding Phase}\label{s:tniep}

As shown by Takizawa (1999), electron temperature distribution is 
significantly different than the plasma mean one in the expanding phase
of mergers. Let us examine two-temperature nature of ICM for each model.

\begin{figure}
    \epsfxsize = 12 cm
    \centerline{\epsfbox{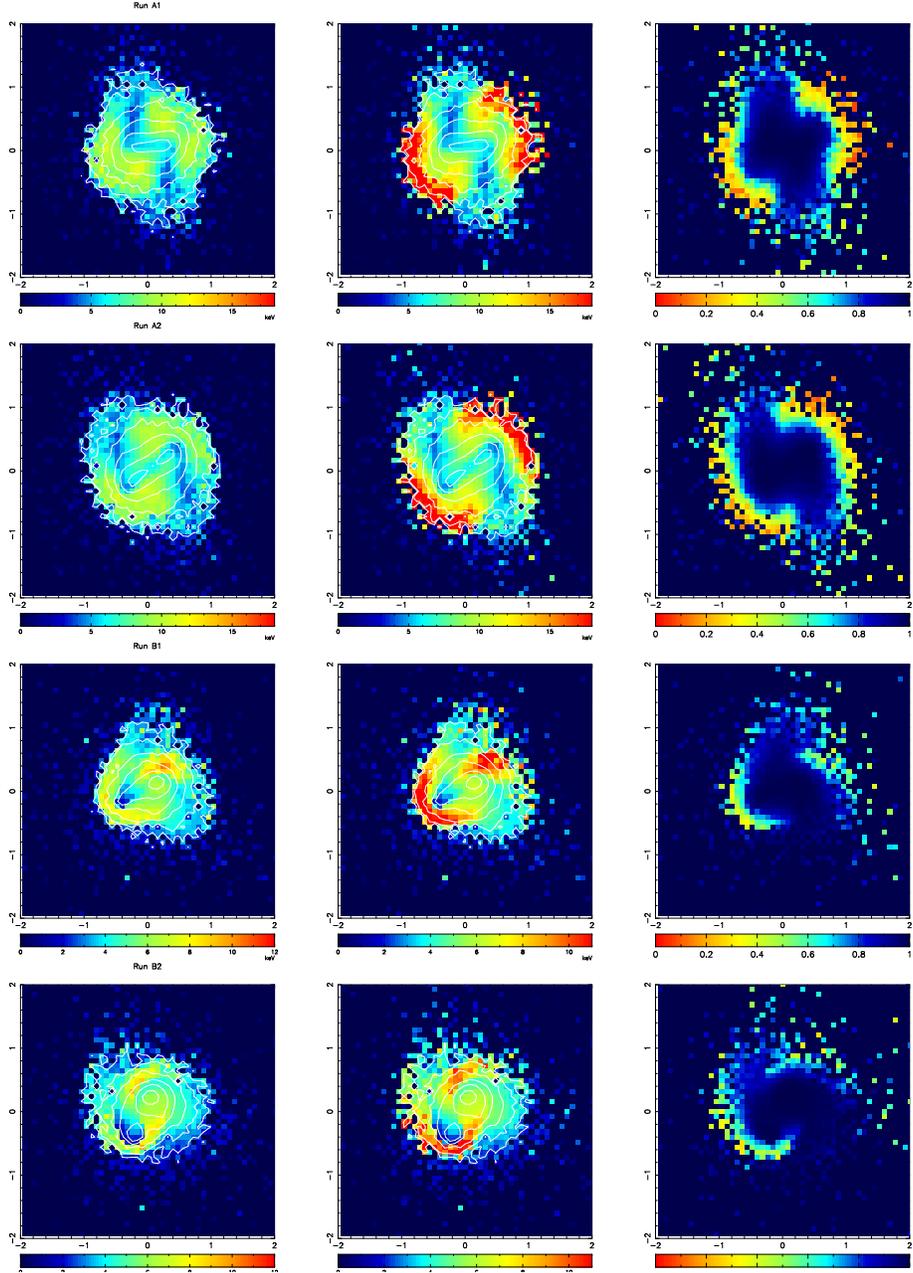}}
\caption{The snapshots of electron temperature ($T_{\rm e}$) 
         distribution (Column 1), the plasma mean temperature ($\bar{T}$) 
         one (Column 2), and the normalized electron temperature 
        ($\tilde{T}_{\rm e} \equiv T_{\rm e}/\bar{T}$) one (Column 3) 
         for each model at expanding phase (0.25 Gyr after the most contracting
         epoch) viewed along the z-axis. X-ray surface 
         brightness distribution (contours) is overlaid for the panel 
         in Column 1 and 2. Run A1, A2, B1, and B2 are showed in Row 1, 
         Row 2, Row 3, and Row 4, respectively. For $\tilde{T}_{\rm e}$ 
         distribution (Column 3), the red, yellow. green, and blue colors 
         correspond to $\tilde{T}_{\rm e} \sim 0.1, 0.3, 0.5$, and $0.7$, 
         respectively.}
\label{fig:twtem}
\end{figure}

Figure \ref{fig:twtem} shows the snapshots of electron temperature 
($T_{\rm e}$) distribution (Column 1), the plasma mean temperature 
($\bar{T}$) one (Column 2), and the normalized electron temperature 
($\tilde{T}_{\rm e} \equiv T_{\rm e}/\bar{T}$) one (Column 3) 
for each model at the expanding phase 
(0.25 Gyr after the most contracting epoch)
viewed along the z-axis. X-ray surface 
brightness distribution (contours) is overlaid for the panels in Column 1 
and 2. Run A1, A2, B1, and B2 are showed in Row 1, Row 2, Row 3, and 
Row 4, respectively. Temperature color scale for $T_{\rm e}$ and $\bar{T}$ 
(Column 1 and 2) in Row 1 and 2 (for Run A1 and A2) , and Row 3 and 4 
(for Run B1 and B2) are the same as in figure \ref{fig:xrsbtea1} and 
\ref{fig:xrsbteb1}, respectively. For $\tilde{T}_{\rm e}$ distribution 
(Column 3), the red, yellow. green, and blue colors correspond to 
$\tilde{T}_{\rm e} \sim 0.1, 0.3, 0.5$, and $0.7$, respectively.

For all models, spiral-shaped shocks are recognized. As in head-on 
mergers (Takizawa 1999), electron temperature in the post shock regions 
is significantly lower than the plasma mean one in this phase. 
However, even in the electron temperature distribution, spirally high
temperature regions can be seen. Note that these regions are located
not at the shocks, but $\sim 0.5$ Mpc behind the shocks.

In the higher angular momentum cases (Run A2 and B2), double-peak
structure can be seen even in the expanding phase whereas it can
be seen only in the pre-merger contracting phase in head-on collisions.
Thus, a cluster which has double peaks in X-ray surface brightness 
distribution and hot temperature regions not between the two peaks
but around them, is most likely a candidate of a off-center collision. 
This is the case of A3395 (Markevitch et al. 1998). This will be discussed
in \S \ref{s:sd}.

\subsection{Line-of-sight Velocity Distribution of Galaxies}\label{s:lvdg}

As described in \S \ref{s:xietmra1} and \ref{s:xietmrb1}, when we observe
merging clusters nearly along the collision axis, line-of-sight (LOS)
velocity distribution of galaxies has more useful information than 
the X-ray images. Let us discuss this issue. In the argument below, 
we assume that DM particles distribution in velocity space well represents 
that of galaxies.

\begin{figure}
    \epsfxsize = 12 cm
    \centerline{\epsfbox{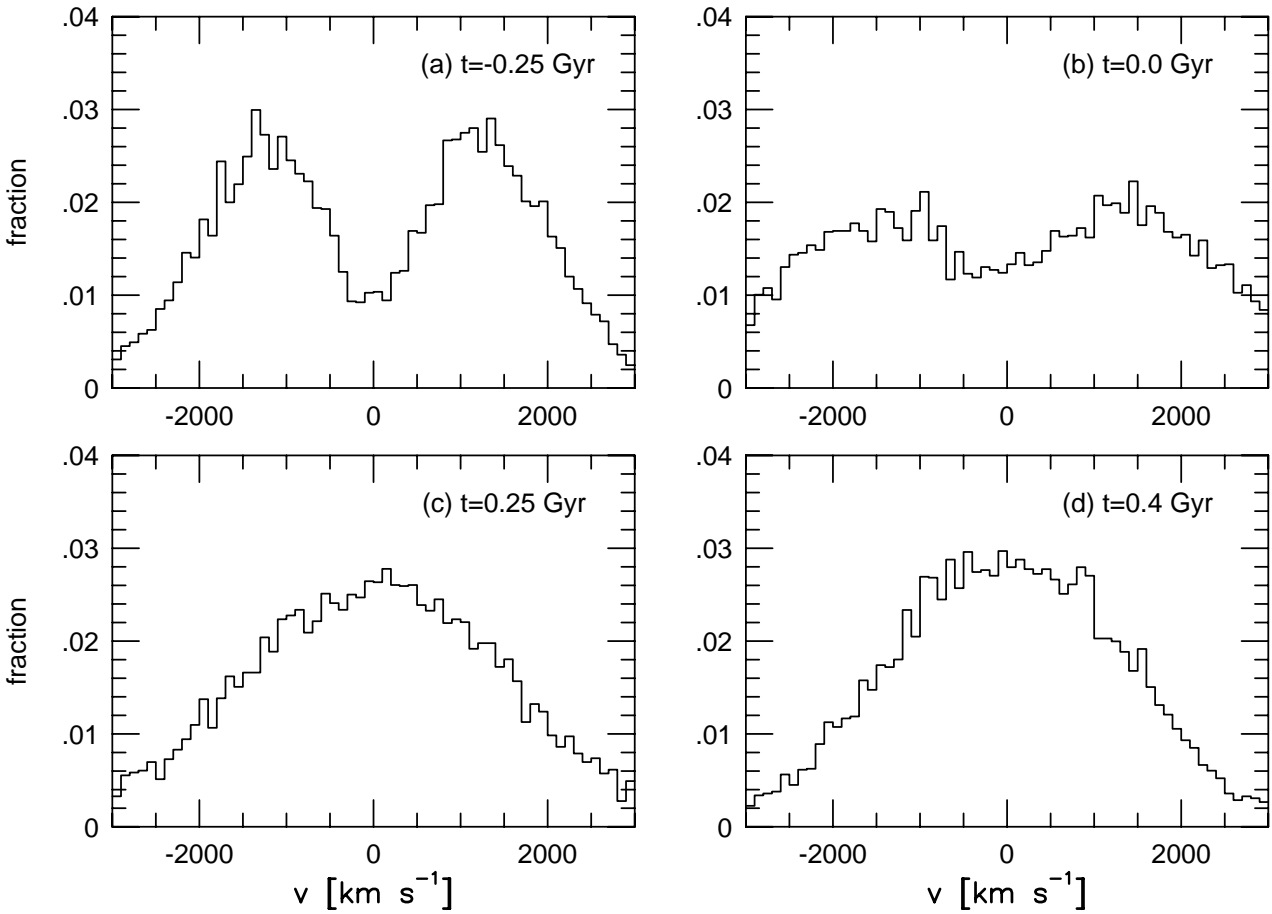}}
\caption{The histograms of line-of-sight velocities of N-body particles of 
         Run A1 viewed along the x-axis.figure caption}
\label{fig:vhga1}
\end{figure}

\begin{figure}
    \epsfxsize = 12 cm
    \centerline{\epsfbox{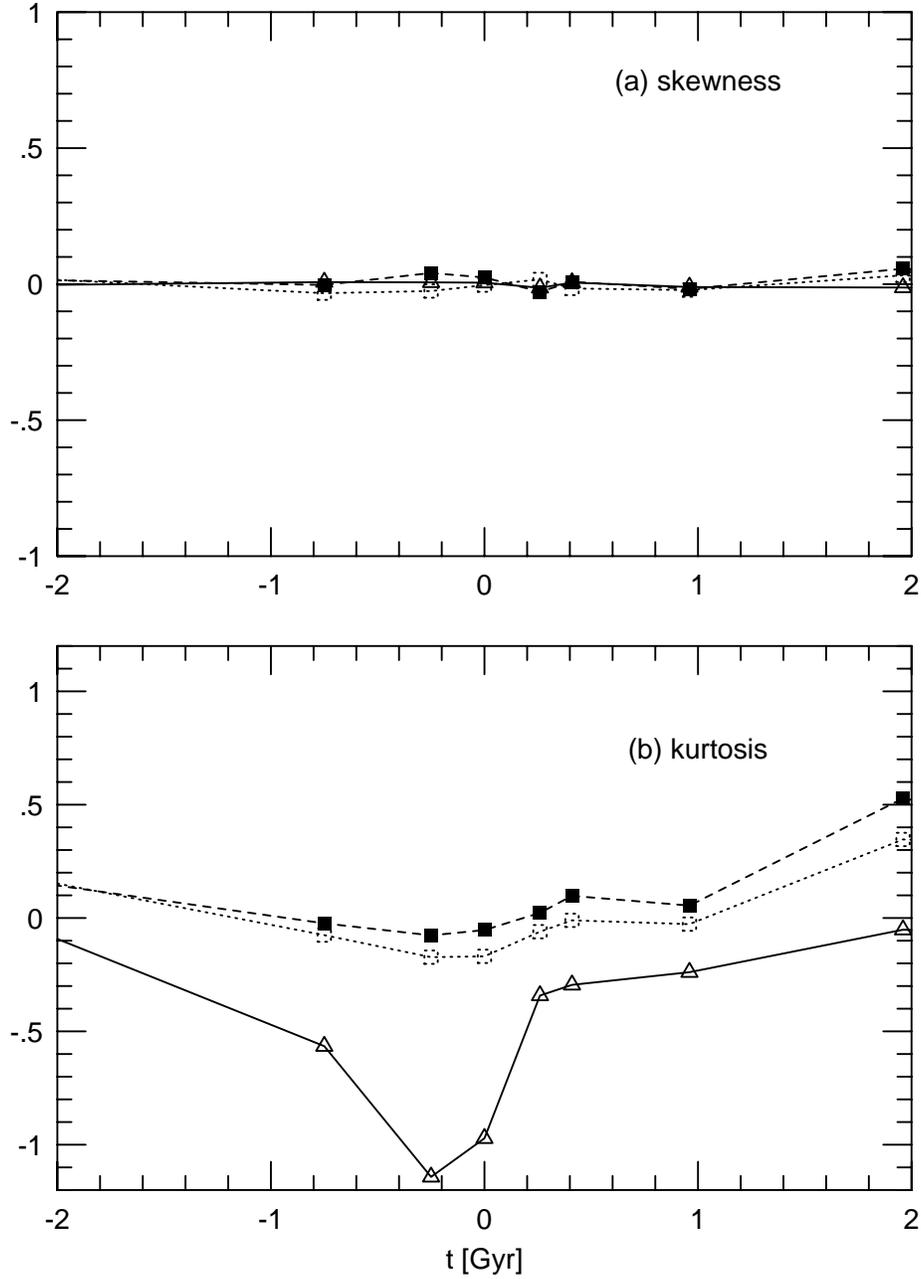}}
\caption{The evolution of the skewness and kurtosis of the distribution
         of line-of-sight velocities along the x-axis 
         (open triangles and solid lines), y-axis (open squares and dotted 
          lines), and z-axis (filled square and short dashed lines), 
          respectively. }
\label{fig:vmoma1}
\end{figure}

First, we describe the evolution of LOS velocity distribution of galaxies 
in Run A1. Figure \ref{fig:vhga1} shows the histograms of
LOS velocities of N-body particles of Run A1 viewed along the 
x-axis at four epochs. For the pre-merger phase ($t=-0.25$ Gyr), 
two components associated
to the subclusters can be clearly recognized. Then at the most contracting
epoch the distribution becomes boxier than the Gaussian ($t=0.0$ Gyr). 
This is confirmed from
the evolution of the kurtosis (Fig. \ref{fig:vmoma1}). However, in the 
expanding phase, the distribution becomes closer to Gaussian. 
To see the evolution of the distribution quantitatively, we show the
evolution of the skewness and kurtosis of the LOS velocity distribution
viewed along the x-axis (open triangles and solid lines), 
y-axis (open squares and 
dotted lines), and z-axis (filled square and short dashed lines), 
respectively. Due to the symmetry of the equal mass collision, it is quite
natural that the skewness is almost zero in all directions. 
On the other hand, the kurtosis of the velocity distribution along the x-axis 
becomes significant negative values especially in the contracting phase 
although the kurtosis of the other components is almost nearly zero.

\begin{figure}
    \epsfxsize = 12 cm
    \centerline{\epsfbox{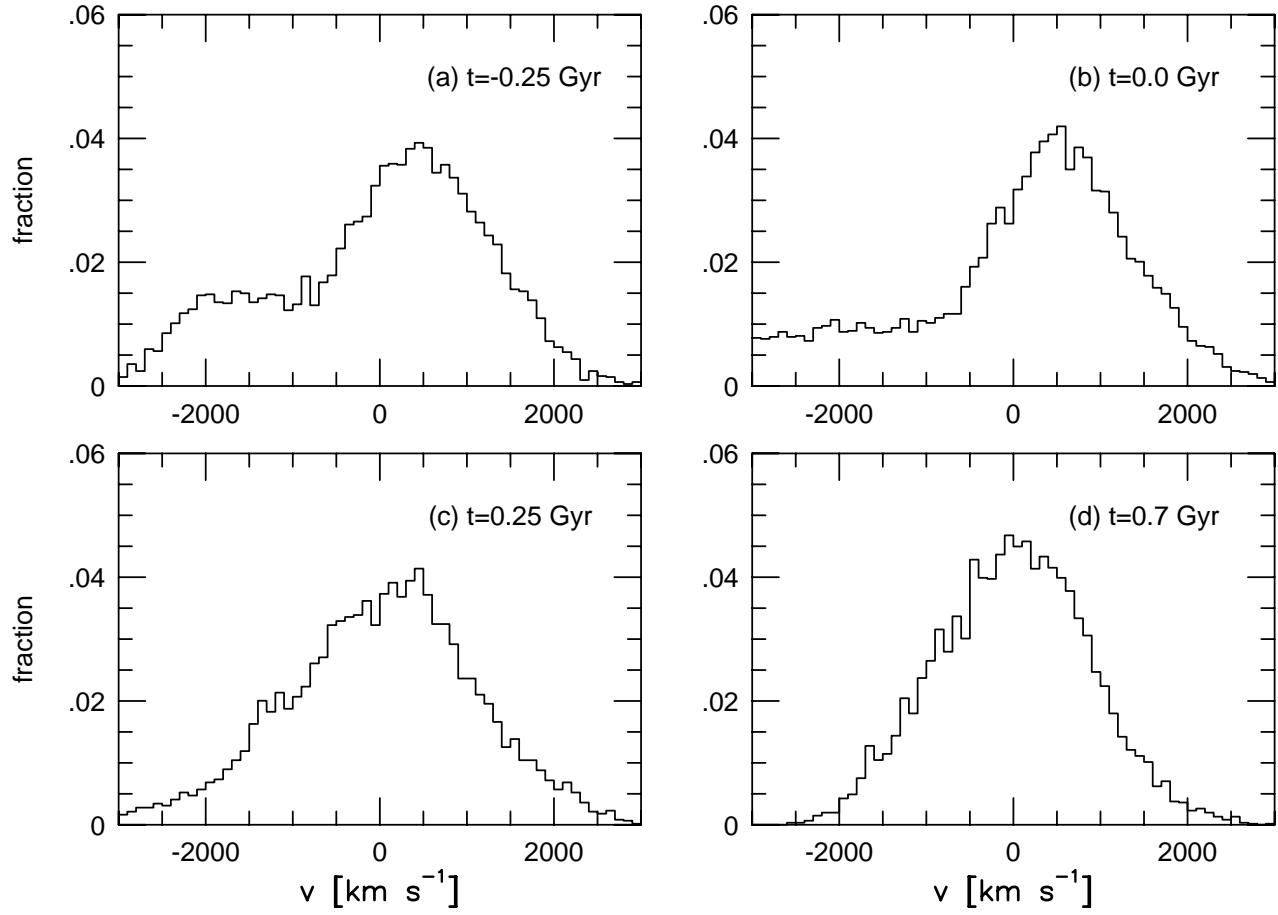}}
\caption{The same as figure \ref{fig:vhga1} but for Run B1.}
\label{fig:vhgb1}
\end{figure}

\begin{figure}
    \epsfxsize = 12 cm
    \centerline{\epsfbox{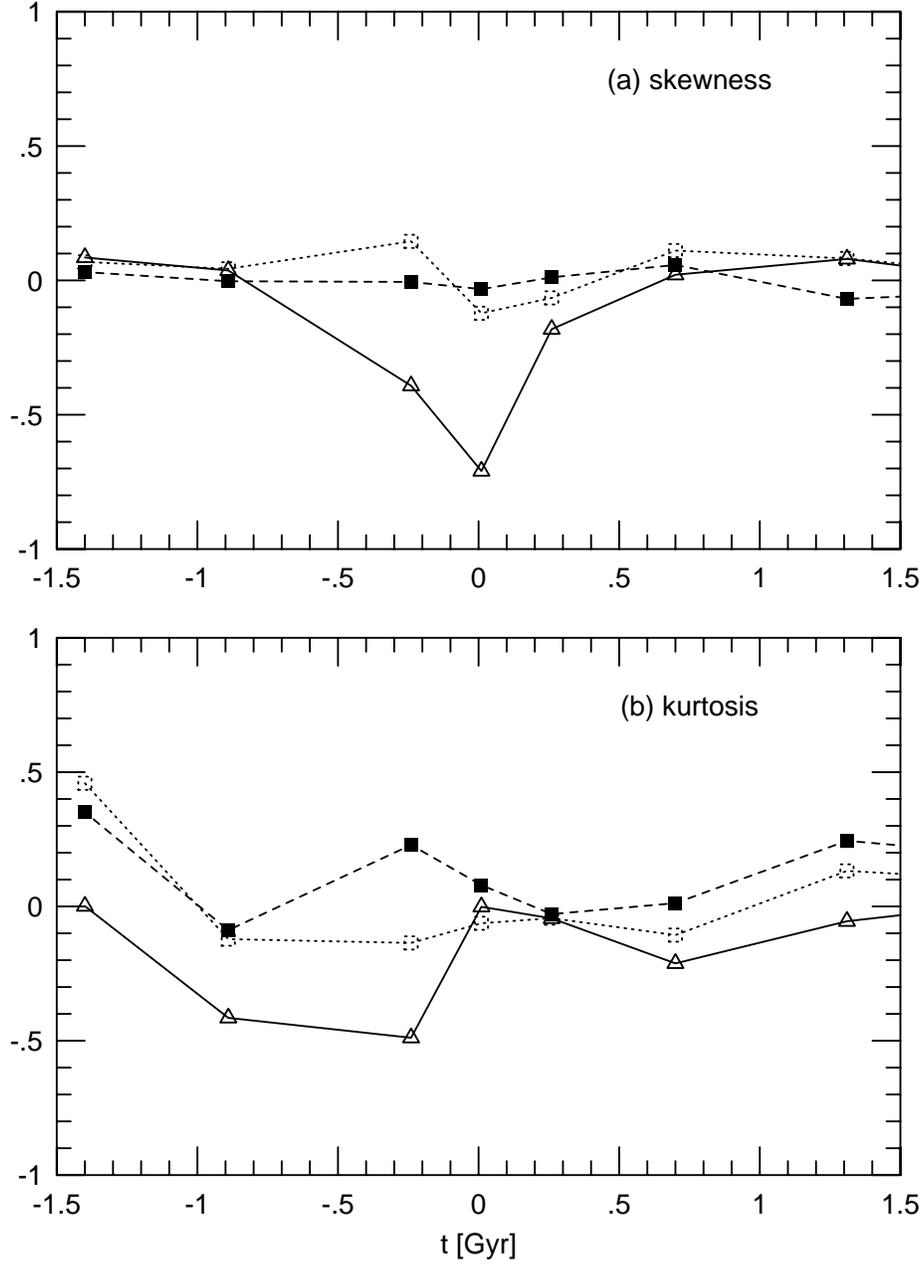}}
\caption{The same as figure \ref{fig:vmoma1} but for Run B1.}
\label{fig:vmomb1}
\end{figure}

Next, we investigate the behavior of Run B1. Figure \ref{fig:vhgb1}
shows the same conditions as figure \ref{fig:vhga1} but for Run B1. 
In the contracting 
phase, a tail associated to the smaller subcluster is visible toward 
negative velocities. This is more clarified in the skewness evolution
in figure \ref{fig:vmomb1}, which is the same conditions as figure 
\ref{fig:vmoma1}
but for Run B1. Due to this tail component the skewness becomes significantly
large negative values. Note that the sign of the skewness depends on the 
relative position among two subclusters and the observer. Again, a kurtosis
of the velocities along the x-axis becomes negative in the contracting phase
although this is not so prominent as in Run A1.

In the above cases, we use the all N-body particles ($N=10000$ for Run A1 and
$N=6250$ for Run B1) to calculating the moments of LOS velocity distribution.
In real cases, however, it is difficult to measure LOS velocities for
such numbers of galaxies. On the other hand, the standard 
deviations of a skewness and a kurtosis for a Gaussian distribution are
approximately $\sqrt{15/N}$ and $\sqrt{96/N}$, respectively.
Thus, for Run A1, we need $\sim 1000$ galaxies' LOS velocities to obtain
a statistically significant negative kurtosis. 
If the number of galaxies
is only $\sim 100$, it is difficult to distinguish a negative kurtosis due to
mergers from that due to poor statistics. For Run B1, $N \simeq 5000$ is 
required to detect a statistically significant negative kurtosis. 
On the other hand, we need $\sim 1000$ galaxies to obtain a 
statistically significant skewness. This will be discussed in \S \ref{s:sd}.

\section{SUMMARY AND DISCUSSION}\label{s:sd}

We investigate evolution of CGs and their observational consequences
during off-center mergers explicitly
considering the relaxation process between the ions and electrons. 
In the contracting phase, a bow shock occurs between 
the two subclusters. X-ray images have two peaks and the high temperature 
region is located between the two. However, the observed temperature 
between the two peaks depends on the viewing angle. In the expanding phase
spiral-shaped two shocks propagate outward. 
Spatial distribution of ICM electron temperature is significantly different
in this phase as in head-on collisions.
Morphological features in X-ray images and temperature distribution 
characteristic to off-center mergers 
are seriously affected by the viewing angle. When we observe CGs during 
off-center merger in some directions, we cannot distinguish them from
head-on mergers only through X-ray images and temperature maps.

When we observe merging CGs nearly along the collision axis, LOS velocity 
distribution of galaxies has more useful information than X-ray images and 
temperature maps of ICM. In the contracting phase, the distribution is boxier
and have a negative kurtosis. This is prominent when the two subclusters have 
nearly equal masses. When the two subclusters have rather different masses,
the skewness of the distribution is more suitable for a indicator
of mergers in the contracting phase. To obtain statistically 
significant values of these moments, we need $\sim 1000$ galaxies' LOS 
velocities. 

Recently, Markevitch et al. (1999) and Kikuchi et al. (1999) estimated
subcluster collision velocities in Cygnus A and Virgo using the electron
temperature maps obtained by ASCA, respectively. Both of them have simple
geometry; Hot temperature region is located just between the two peaks
of X-ray images. However, if we do not observe CGs just perpendicular 
to the collision axis, the observed temperature between the peaks is 
lower than the true temperature at the shock. This is due to the
contamination of cooler ICM located foreground and background. Therefor,
this method can lead to underestimation of the collision velocities. 
We should regard their results as the minimum limit. 
However, since estimated velocities through electron temperature maps 
depend on the true collision velocities and the viewing angles,  
they are related to the transverse components of the collision
velocities. On the other hand, LOS components of collision velocities
will be observable with XRS after the launch of ASTRO-E. 
Therefor, we will be able to estimate true collision velocities
combining the results of both electron temperature maps and LOS velocities.

When we observe merging clusters around the most contracting phase nearly
along the collision axis through X-ray, they seem to be relaxed spherical 
ones in X-ray images. In addition, the $\beta_{\rm spec}$ is nearly unity
only around this epoch though a large $\beta_{\rm spec}$ value is a good 
indicator of mergers when observed along the collision axis
(Ishizaka 1997; Takizawa 1999). Thus, it is difficult to resolve whether 
they are merging 
clusters or not. One of their characteristic features is a large temperature
gradient. Markevitch (1996) found that some of clusters have a larger
radial gradient in electron temperature than that expected from 
the self-similar solution (Bertschinger 1985) and numerical simulations 
(eg; Navarro et al. 1995; Takizawa \& Mineshige 1998) in the plasma mean
temperature. One explanation to solve
this discrepancy is energy nonequipartition of electrons and ions in the 
outer regions (Fox \& Loeb 1997; Chi\'eze, Alimi \& Teyssier 1998; 
Ettori \& Fabian 1998; Takizawa 1998). However, it is probable that
some of them are hidden merging clusters as mentioned above.
Again, measurement of ICM LOS velocities with XRS of ASTRO-E will clarify
such hidden merging clusters.

When the angular momentum of the systems are relatively high 
($\lambda \sim 0.02$), double-peak structures in X-ray images survive
after the most contracting epoch. We cannot find whether they are in 
a contracting or expanding phase only through X-ray surface brightness 
distribution. However, electron temperature distribution enable
us to determine the phase of mergers. If high temperature regions are
located not between but around the peaks like A3395 
(Markevitch et al. 1998), they are in the expanding phase of off-center 
mergers. If high temperature regions are between the peaks like Cygnus A 
(Markevitch et al. 1999), on the other hand, they are in contracting phase.

If we assume the luminosity function of galaxies is the 
Schechter form (see Binney \& Merrifield 1998) 
with the normalization through a typical galaxy number 
density in rich clusters, the absolute $B$ magnitude of the 1000-th luminous
galaxy is $\sim -17$. This corresponds to $19.5$ in the apparent 
$B$ magnitude for an object at $z=0.05$ when the Hubble constant is 
75 km ${\rm s}^{-1}$ ${\rm Mpc}^{-1}$.
This is nearly equal to the survey limit of 2dF ($B=19.5$).
Thus, we will be able to discuss on merger dynamics for nearby clusters
using such redshift survey data.

\acknowledgements
The author would like to thank K. Ohta, T. T. Takeuchi, and T. T. Ishii
for valuable discussion on the observational feasibility of 
galaxies' LOS velocities in clusters.
The author also thanks K. Masai and N. Y. Yamasaki for helpful comments and  
S. Mineshige for continuous encouragement.

\clearpage
 \begin{table}
  \begin{center}
   \begin{tabular}{ccccc} 
    \hline \hline
                             & Run A1 & Run A2 & Run B1 & Run B2 \\ 
    \hline
$M_1/M_2 (10^{15} M_{\odot}$)& 0.5/0.5 & 0.5/0.5 & 0.5/0.125 & 0.5/0.125  \\
$r_{{\rm c},1}/r_{{\rm c},2}$  (Mpc)& 0.2/0.2 & 0.2/0.2 & 0.2/0.1 & 0.2/0.1 \\
$k_{\rm B}T_1/k_{\rm B}T_2$ (keV)& 4.78/4.78 & 4.78/4.78 & 4.78/2.39 & 4.78/2.39 \\
$R$ (Mpc)                    & 3.2 & 3.2 & 2.2 & 2.2 \\
$b$ (Mpc)                    & 0.4648 & 1.042 & 0.5394 & 1.456 \\
$v_{\rm init}$ (km/sec)      & 714 & 637 & 863 & 639 \\
$\lambda$                    & 0.01 & 0.02 & 0.01 & 0.02  \\
$f_{\rm g}$                  & 0.1 & 0.1 & 0.1 & 0.1 \\
$\epsilon$ (Mpc)             & 0.02 & 0.02 & 0.01 & 0.01 \\
$N_1/N_2$ (SPH)              & 5000/5000 & 5000/5000 & 5000/1250 & 5000/1250      \\
$N_1/N_2$ (DM)               & 5000/5000 & 5000/5000 & 5000/1250 & 5000/1250      \\
    \hline
   \end{tabular}
  \end{center}
  \caption[Model parameters]{Model parameters}
  \label{tab:para}
 \end{table}%

\clearpage
\begin{references}
\reference{} Barnes, J., \& Hut, P. 1986, Nature, 324, 446
\reference{} Barnes, J., \& Efstathiou, G. 1987, ApJ, 319, 575
\reference{} Bertschinger, E. 1985, ApJS, 58, 39
\reference{} Bekki, K. 1999, ApJ, 510, L15
\reference{} Binney, J., \& Merrifield, M. 1998, Galactic Astronomy
             (Princeton: Princeton University Press)
\reference{} Binney, J., \& Tremaine, S. 1987, Galactic Dynamics 
             (Princeton: Princeton University Press)
\reference{} Bryan, G. L., \& Norman, M. L., 1998, ApJ, 495, 80
\reference{} Caldwell, H., Rose, J. A., Sharples, R. M., Ellis, R. S.,
             \& Bower, R. G. 1993, AJ, 106, 473
\reference{} Chi\'eze, J.-P., Alimi, J.-M, \& Teyssier, R. 1998, ApJ, 495, 630
\reference{} Churazov, E., Gilfanov, M., Forman, W., \& Jones, C. 1998, 
             ApJ, submitted (astro-ph/9802166)
\reference{} Deiss, B. M., Reich, W., Lesch, H., Wielebinski, R. 
             1997, A\&A, 321, 55
\reference{} Donnelly, R. H., Markevitch, M., Forman, W., Jones, C., 
             David, L. P., Churazov, E., \& Gilfanov, M. 1998, ApJ, 500, 138
\reference{} Donnelly, R. H., Markevitch, M., Forman, W., Jones, C., 
             Churazov, E., \& Gilfanov, M. 1999, ApJ, 513, 690
\reference{} Eke, V. R., Navarro, J. F., \& Frenk, C. S. 1998, ApJ, 503, 569
\reference{} Ettori, S. \& Fabian, A. C. 1998, MNRAS, 293, L33
\reference{} Evrard, A. E. 1990, ApJ, 363, 349
\reference{} Evrard, A. E. 1991, MNRAS, 248, 8
\reference{} Fox, D. C., \& Loeb, A. 1997, ApJ, 491, 459
\reference{} Fujita, Y., Takizawa, M., Nagashima, M., \& Enoki, M. 1999, PASJ, 
             in press (astro-ph/9904386)
\reference{} Fusco-Femiano, R., Fiume, D. D., Feretti, L., Giovannini, G.,
             Grandi, P., Matt, G., Molendi, S., and Santangelo, A.
             1999, ApJ, 513, L21
\reference{} Giovannini, G., Feretti, L., Venturi, T., Kim, K. T., and
             Kronberg, P. P. 1993, ApJ, 406, 399
\reference{} Henriksen, M. J., \& Markevitch, M. L. 1996, 466, L79
\reference{} Hernquist, L., \& Katz, N. 1989, ApJS, 70, 419
\reference{} Honda, H., Hirayama, M., Watanabe, M., Kunieda, H., Tawara, Y., 
             Yamashita, K., Ohashi, T., Hughs, J., P., \& Henry, J. P. 
             1996, ApJ, 473, L71
\reference{} Ishizaka, C., \& Mineshige, S. 1996, PASJ, 48, L37
\reference{} Ishizaka, C., 1997 Ap \& SpS, 254, 233
\reference{} Hanami, H., Tsuru, T., Shimasaku, K., Yamauchi, S., Ikebe, Y., 
             \& Koyama, K. 1999, ApJ, in press
\reference{} Kikuchi, K. et al. 1999, in preparation
\reference{} Markevitch, M. 1996, ApJ, 465, L1
\reference{} Markevitch, M., Forman, W. R., Sarazin, C. L., \& Vikhlinin, A.
             1998, ApJ, 503, 77
\reference{} Markevitch, M., Sarazin, C. L., \& Vikhlinin, A. 1999, ApJ, 
             in press (astro-ph/9812005)
\reference{} Monaghan, J. J. 1992, ARA\&A, 30, 543
\reference{} Navarro, J. F., Frenk, C. S., \& White, S. D. M. 1995, 
             MNRAS, 275, 720 
\reference{} Ricker, P., M. 1998, ApJ, 496, 670
\reference{} Roettiger, K., Burns, J. O., \& Loken, C. 1996, ApJ, 473, 651
\reference{} Roettiger, K., Stone, J. M., \& Burns, J. O.
             1999, ApJ, in press (astro-ph/9902015)
\reference{} R\"{o}ttgering, H. J. A., Wieringa, M. H., Hunstead, R. W., 
             and Ekers, R. D. 1997, MNRAS, 290, 577
\reference{} Sarazin, C. L. 1999, ApJ, in press (astro-ph/9901601)
\reference{} Schindler, S., \& M\"{u}ller, E. 1993, A\&A, 272, 137
\reference{} Suginohara, T., \& Ostriker, J. P. 1998, ApJ, 507, 16
\reference{} Takizawa, M., \& Mineshige, S. 1998, ApJ, 499, 82
\reference{} Takizawa, M. 1998, ApJ, 509, 579
\reference{} Takizawa, M. 1999, ApJ, in press (astro-ph/9901314)
\reference{} Tomita, A., Nakamura, F. E., Takata, T., Nakanishi, K.,
             Takeuchi, T., Ohta, K., \& Yamada, T.
             1996, AJ, 111, 42
\reference{} Ueda, H., Shimasaku, K., Suginohara, T., \& Suto, Y. 1994, 
             PASJ, 46, 319
\reference{} Wang, Q. D., Ulmer, M. P., Lavery, R. J. 1997, MNRAS, 288, 702
\reference{} Watanabe, M., Yamashita, K., Furuzawa, A., Kunieda, H., 
             Tawara, Y., \& Honda, H. 1999, ApJ, submitted 
\reference{} Watanabe, M., Yamashita, K., \& Furuzawa, A. 1999, in preparation
\reference{} Yoshikawa, K., Itoh, M., \& Suto, Y. 1998, PASJ, 50, 203
\end{references}
\end{document}